\documentclass{article}

\usepackage{microtype}
\usepackage{graphicx}
\usepackage{subfigure}
\usepackage{booktabs} % for professional tables
\usepackage{url}            % simple URL typesetting
\usepackage{amsfonts}       % blackboard math symbols
\usepackage{nicefrac}       % compact symbols for 1/2, etc.
\usepackage{microtype}      % microtypography
\usepackage{xcolor}         % define colors in text
\usepackage{xspace} 
\usepackage{amsmath}
\usepackage{amssymb}
\usepackage{ctable}
\usepackage{multirow}
\usepackage{hhline}

\newtheorem{definition}{Definition}
\newtheorem{theorem}{Theorem}

\newtheorem{lemma}{Lemma}

\usepackage{hyperref}

\usepackage[accepted]{icml2021}

\icmltitlerunning{IRLI}

\begin{document}

\twocolumn[
\icmltitle{IRLI: Iterative Re-partitioning for Learning to Index}



\icmlsetsymbol{equal}{*}

\begin{icmlauthorlist}
\icmlauthor{Gaurav Gupta}{equal,ece}
\icmlauthor{Tharun Medini}{equal,ece}
\icmlauthor{Anshumali Shrivastava}{cs}
\icmlauthor{Alexander J Smola}{amz}
% \icmlauthor{Fiuea Rrrr}{to}
% \icmlauthor{Tateu H.~Yasehe}{ed,to,goo}
% \icmlauthor{Aaoeu Iasoh}{goo}
% \icmlauthor{Buiui Eueu}{ed}
% \icmlauthor{Aeuia Zzzz}{ed}
% \icmlauthor{Bieea C.~Yyyy}{to,goo}
% \icmlauthor{Teoau Xxxx}{ed}
% \icmlauthor{Eee Pppp}{ed}
\end{icmlauthorlist}

\icmlaffiliation{ece}{Department of Electrical Engineering, Rice University, Houston, Texas}
\icmlaffiliation{cs}{Department of Computer Science, Rice University, Houston, Texas}
\icmlaffiliation{amz}{Amazon Web Services, Palo Alto, California}

\icmlcorrespondingauthor{Gaurav Gupta}{gaurav.gupta@rice.edu}
\icmlcorrespondingauthor{Tharun Medini}{tharun.medini@rice.edu}

\icmlkeywords{Machine Learning, ICML}

\vskip 0.3in
]



\printAffiliationsAndNotice{\icmlEqualContribution} % otherwise use the standard text.

\begin{abstract}

Neural models have transformed the fundamental information retrieval problem of mapping a query to a giant set of items. However, the need for efficient and low latency inference forces the community to reconsider efficient approximate near-neighbor search in the item space. To this end, learning to index is gaining much interest in recent times. Methods have to trade between obtaining high accuracy while maintaining load balance and scalability in distributed settings. We propose a novel approach called IRLI (pronounced \textit{`early'}), which iteratively partitions the items by learning the relevant buckets directly from the query-item relevance data. Furthermore, IRLI employs a superior power-of-$k$-choices based load balancing strategy. We mathematically show that IRLI retrieves the correct item with high probability under very natural assumptions and provides superior load balancing. IRLI surpasses the best baseline's precision on multi-label classification while being $5x$ faster on inference. For near-neighbor search tasks, the same method outperforms the state-of-the-art Learned Hashing approach NeuralLSH by requiring only $\approx \frac{1}{6}^{th}$ of the candidates for the same recall. IRLI is both data and model parallel, making it ideal for distributed GPU implementation. We demonstrate this advantage by indexing 100 million dense vectors and surpassing the popular FAISS library by $>10\%$ on recall.

\end{abstract}

\section{Introduction}
% The K-nearest neighborhood (KNN) has been widely studied in Machine learning and Information retrieval community
For a given query $q$, the classical problem in information retrieval (IR) is to learn a function $f$ that maps $q$ to one (or few) of an extensive set (often hundreds of millions) of discrete item sets.  Most practical applications in IR have data with only the query-item relevance (like the classical query-product purchase data \cite{dssm,mach}). Modern search engines often deploy a pipeline in which queries and items are first embedded into a dense vector space. Here the information retrieval problem gets reduced to the Approximate Near Neighbor (ANN) search in the embedding space. Analogs of this approach are explored in industry-scale works like SLICE~\cite{jain2019slice} (Bing Search) and DSSM~\cite{dssm} (Amazon Search).

Approximate Near Neighbor (ANN), being one of the most studied algorithms in machine learning, is still far from being solved to a satisfactory extent in the context of information retrieval. In the past decade, learning-based solution for ANN has shown significant promise. Surprisingly, learning-based ANN is precisely the same information retrieval problem of finding the function $f$ that maps $q$ to one (or few) of an extensive set (often hundreds of millions) of discrete neighbors. The hardness of the fact that we are dealing with a large item set remains the primary challenge.  

Learning to index has been a sought-after solution to the fundamental challenge of large item space, particularly in the context of Approximate Near Neighbor (ANN) search~\cite{kraska2018case}. The idea behind learning to index is to find a function $f$ that maps the query $q$ to a reasonably sized $B$ discrete partitions, where $B$ is much smaller than the number of total items $L$.  The hope is that if the partitions are reasonably balanced, then the function $f$ reduces the search space from $L$ to $\frac{L}{B}$, which is manageable for large enough $B$.

Several approximate algorithms have been proposed for learning to index. These algorithms reduce the search space, mainly using space partitioning~\cite{nlsh} or graph-based methods~\cite{malkov2018efficient}, or by reducing the complexity of distance computation such as quantization and lookup table based methods~\cite{faiss}. Graph-based NNS methods are efficient but limited to small-scale datasets. Due to their sequential nature, it is not trivial to parallelize the query and indexing process. Given this, many commercial applications that deal with large-scale data increasingly use partitioning based approaches.

Overall, the current IR pipeline still struggles with two significant challenges - 1) Learning embedding from query-item relevance is a pairwise training process leading to a massive amount of training samples and extended training time. Additionally, negative sampling techniques have to be employed to prevent degenerate solutions, which would only exacerbate the problem for large output spaces. 2) The power-law distribution of data causes an imbalanced partition of items into the hash buckets. As a result, frequently queried items tend to coalesce in large numbers into a few buckets, leaving the infrequent ones in the remaining buckets. This imbalance leads to higher inference times as we query heavy buckets more often than the lighter ones. 
\begin{figure*}[ht]
\vskip 0.2in
\begin{center}
\centerline{\includegraphics[scale=0.68]{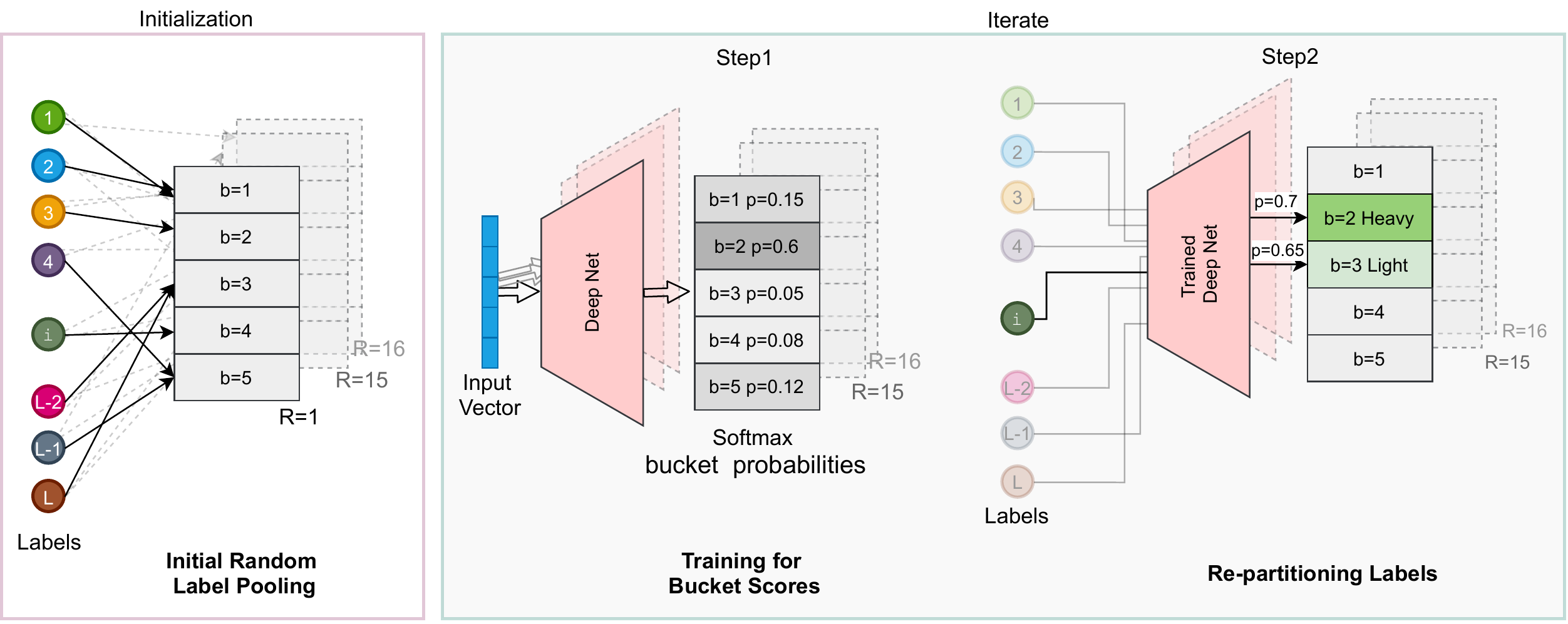}}
\vspace{-0.2cm}
\noindent\caption{We create IRLI index in 3 steps. First \textbf{(left)} (initialization step)- the labels are pooled randomly into $B$ buckets using a 2-universal hash function. The above figure shows only five buckets (while we have a few thousands in practice). Second \textbf{(middle)}- We train $R$ fully-connected networks on $N$ data points, where any bucket containing at-least one true labels is positive. Third \textbf{(right)}: After training for a few epochs, the labels are re-assigned to the buckets. For each label, we provide a representative input to the $R$ networks. We select the top-$K$ buckets and assign the label to the least occupied bucket (K=2 in the figure yields $2^{nd}$ and $3^{rd}$ buckets as the top-scored ones. Light-green bucket is the lesser occupied one, and hence we assign the label to the $3^{rd}$ bucket). A larger $K$ ensures perfect load balance, while a smaller $K$ ensures higher precision and recall.}
\label{fig:NeuralIndexCreation}
\end{center}
\vskip -0.2in
\end{figure*}

\subsection{Our contributions}
\noindent\textbf{1)} We propose a learning-to-index algorithm - IRLI, which learns to partition and map together using a single neural network via an alternative training and re-partitioning.\\ 
\textbf{2)} We theoretically show that IRLI achieves superior recall while maintaining load-balance in convergence.\\
\textbf{3)} We surpass the best Learned Indexing baselines on recall by probing $<20\%$ candidates. We outperform the best Extreme Classification baselines on precision while inferring $5 \times$ faster.\\
\textbf{4)} We index a 100MM sample of the Deep-1B dataset by a trivial distribution of the vectors across eight nodes. With a 4 ms latency on CPU, we achieve a recall of $90\%$, which is a good $10\%$ over the popular FAISS library.

\section{Related works}
There are several well-known approaches for partitioning like k-means clustering \cite{wang2011improved}, locality sensitive hashing (LSH) \cite{shrivastava2014densifying,crossPlotopLSH,multiProbeLSH} and tree-based methods like PCA trees \cite{pcaTrees} and Randomised Partition trees \cite{RPTrees}. LSH in particular is a cheap and fast solution for high-dimensional data indexing. However, both LSH and k-means generate partitions that have extremely skewed load for lop-sided distributed data. The tree-based methods on the other hand suffer from the curse of dimensionality. Additionally, their hierarchical design diminishes their parallelism during the query and construction process.

The power law distribution issue in traditional indexing has warranted research in load-balanced indexing schemes. An early attempt in this regard is Balanced K-Means \cite{malinen2014balanced}, which has high construction time dampening its scalability. A recent noteworthy work is \textbf{NeuralLSH} (NLSH) \cite{nlsh}, which uses KaHIP \cite{KaHIP}, a balanced graph-partitioning algorithm. NLSH maps a query to the relevant partition/partitions \textit{via} a brute force approach, using the partition-centers' distance with the query. However, the centroids do not always reflect the higher-order statistics of the partitions. Sometimes, they do get drifted by outliers within the paritions. NLSH learns a model to rank the partitions generated by the $k$-NN graph. Learning improves the mapping by training on the true query to partition affinity.\\
\textbf{Parabel:} In the context of Information Retrieval, Parabel~\cite{parabel} is one of the primary algorithms that partitions the label space into roughly equal sets via a balanced 2-means label tree, where the label vectors are constructed using input instances. Subsequent improvements like eXtremeText~\cite{extremeText} and Bonsai~\cite{bonsai} relax the 2-way partitioning to higher orders of hierarchy.

\textbf{SLICE:} Another recent work, SLICE, builds an ANN graph on the label vectors obtained from a pre-trained network. It maps a query to the common embedding space during inference and performs a random walk on the ANN graph to obtain the relevant labels.

All existing approaches decouple the partitioning step from the learning step. Once a partition is created, it is fixed for the rest of the process while we map the query using either centroids, hashing, or a learned model. In many cases, the partitioning process is an off-the-shelf algorithm (like KaHIP). Our work differs from the prior work primarily in the fact that we alternatively learn both the mapping and partitions. We have a single model that maps the query to buckets and also partitions the labels for subsequent training. To improve the candidate set precision, we repeat this process in few independent repetitions.

\section{Our Method: IRLI}\label{NeuralIndex}
Iterative Re-partitioning for Learning to Index (IRLI) begins with a random-pooling based index initialization followed by an iterative process of alternating train and re-partition steps. We train $R$ independent such indexes and use them for efficient item retrieval. Figures \ref{fig:NeuralIndexCreation} and \ref{fig:NeaurlIndexQuery} illustrate our algorithm with a toy example of $5$ buckets.

\paragraph{Notation:} For a given dataset $\mathcal{D}$, we denote a query vector by $x$ and its labels by $\bar{y}$. Let $N$ be the total number of train vectors, $d$ be the input vector dimension, and $L$ be the total number of labels. For the ANN scenario, $L=N$. $R$ is the number of repetitions (independent indexes), $B$ is the number of partitions in each repetition, and $f(.)$ is the learned deep-net model (we have $R$ such models).

\subsection{Initialization:} We initialise the partitioning randomly. For this, we use $R$ 2-universal hash functions $h_r: [L] \rightarrow [B],\ r \in \{1,2,..,R\}$. The hash function $h_r(.)$ uniformly maps the $L$ labels into $B$ buckets. As the pooling is randomised, the buckets contain an equal number of labels in expectation. If the application is near neighbor, the $N$ vectors gets pooled randomly $R$ number of times into $B$ partitions. The first part of figure \ref{fig:NeuralIndexCreation} shows the initialization.

\subsection{Alternative Training and Re-partitioning:}\label{sec:repartition} 
\paragraph{Training:} We train a feed forward neural network to learn the affinity $f_r$ of a given point $x$ to $B$ buckets where $B \ll L$. We have $R$ independent partitions and thereby $R$ independent neural networks $\{f_r| r\in [1,2,..,R]\}$.
We are effectively solving a classification problem using the binary cross-entropy loss
$$ \mathcal{L}(x,y,B) = -\sum_{b=1}^{B} y_{o,b} \log \left(p_{b}\right) - (1-y_{o,b})\log(1-p_{b})$$
by providing $B$ softmax scores ($p_{b}$) against the ground-truth one hot value ($y_{o,b}$). $y_{o,b}=1$ if there is at-least one true label present in the bucket $b$, else $y_{o,b}=0$.

In Extreme Multi-Label (XML scenario), the true labels are provided with the data. For the ANN datasets, we use the 100 exact near neighbors to a query point (using the corresponding distance metric) as labels. These neighbors have to be generated beforehand.

\begin{algorithm}[tb]
   \caption{IRLI Index Creation }
   \label{alg:neuralIndexCreationAlgo}
\begin{algorithmic}
   \STATE {\bfseries Input:} data $(x_i,y_i) \in \mathbb{R}^{N \times d}$ and labels $l_i \in \{1..L\}$
   \FOR{$r=1$ {\bfseries to} $R$}
   \STATE $b = h_r(l_i)$, $l_i \in \{1..L\}$ \#Initial assignment
   \STATE Bucket($l_i$) = $b$
   \FOR{$epoch=1$ {\bfseries to} $T$}
    \STATE Learn bucket scoring- $f_r(x_i) \in \mathbb{R}^B$
    \FOR{$l_i=1$ {\bfseries to} $L$}
    \STATE $p_{l_i} = g(l_i)$ \#get label affinity 
    \STATE $\mathcal{B}$ = topK($p_{l_i}$) \#choose top K buckets
    \STATE $b =  $argmin Load$(\mathcal{B})$ 
    \STATE Bucket($l_i$) = $b$ \#re-assignment
    \ENDFOR
   \ENDFOR
   \ENDFOR
 \STATE \#Label affinity $p_{l_i} = g(l_i)$ definition
 \IF {$l_i$ == $x_i$}
 \STATE  $p_{l_i} = f(x_i)$ \#function $g(.)$ and $f(.)$ is the same 
 \ELSE 
 \STATE $p_{l_i} = $Sum$ (f(x_i)\ \forall (x_i, y_i)\ s.t.\ l_i \in y_i$) 
 \#gives the effective buckets' probability score
 \ENDIF
\end{algorithmic}
\end{algorithm}

\paragraph{Re-partitioning}: This is the critical step in IRLI as it creates a partition with more relevant label affinity than the current partition. Let us first define label affinity for both the XML and ANN scenarios. The label affinity in the absence of a label vector (XML scenario) is defined as:
\begin{definition}
\label{affnity2}
For a given label $l$ and a network $f(.)$ with $B$ outputs, trained on a dataset $x,y \in \mathcal{D}$, the label affinity is given by
$$P_l = \sum_{\forall l \in y_i} f(x_i), f(.)\in R^B$$
\end{definition}
When the label vector is given (ANN scenario) the label affinity is defined as:
\begin{definition}
\label{affnity1}
For a given label embedding $\overline{l}$ and a network $f(.)$ with $B$ outputs, the label affinity is given by
$$P_l =  f(\overline{l}),  f(.)\in R^B$$
\end{definition}

Please refer to section \ref{analysis} for further analysis on the label affinity behavior.

Once the training is done, a label ($l \in \{1,2..L\}$) is re-partitioned by assigning it to its top affinity bucket ($argmax\ P_l $), in each of the $R$ repetitions. It is important to note that, for similar labels, the network will provide very similar label affinities.

\textbf{Load Balancing}: In general, a real dataset does not have a uniform distribution. Consequently, similarity-based partitioning leads to an unbalanced load across the buckets. To overcome this, we select $K$ top affinity buckets for each label instead of $1$. We choose the least occupied bucket among these top-scoring buckets and assign the label to it. This ensures that the label fills the lighter bucket first to keep up with the load of the top buckets of similar labels. As we observe later in section \ref{experiments}, we will only need a small $K$ to maintain a near-perfect load balance. For example, on GloVe100 dataset, for $B=5000$, we only need $k=10$ buckets to achieve a load variance of approx $3$ as compared to $15$ for random bucket assignment (lower the load-variance, better the balance).

\textbf{Absence of label embedding}: In the case of ANN datasets, label vectors are given beforehand. On the other hand, if a label embedding is not given, we need an equivalent of it to pass through the learned network and get the label affinity. For such cases,  we retrieve the bucket scores for a label using its data points (as shown in definition \ref{affnity2}). For each label $l$, we use the sum of the corresponding data vectors' softmax probability.

We re-assign labels once every few training epochs (once every five epochs in our experiments). We alternate between the training and re-partitioning steps until the number of new assignments converges to zero.

\begin{algorithm}[tb]
   \caption{IRLI Index Query}
   \label{alg:neuralIndexQueryAlgo}
\begin{algorithmic}
   \STATE {\bfseries Input:} Models $f_r(.)$, IRLI $\Pi_r, r \in \{1..R\}$
   \STATE {\bfseries Query point:} $q \in Q$
   \FOR{$r=1$ {\bfseries to} $R$}
    \STATE $\mathcal{B}[1:m,r] = $topm$(f_r(q))$
    \ENDFOR
    \FOR{$\mathbf{b}=\mathcal{B}[1,1]$ {\bfseries to} $\mathcal{B}[m,R]$}
    \STATE $\phi = \phi \cup InvIndex (\mathbf{b})$
    \ENDFOR
    \STATE Candidate set = FrequentOnes$(\phi)$

\end{algorithmic}
\end{algorithm}

\subsection{Inference/Query:} 

\begin{figure}[ht]
\vskip 0.2in
\begin{center}
\centerline{\includegraphics[scale=0.5]{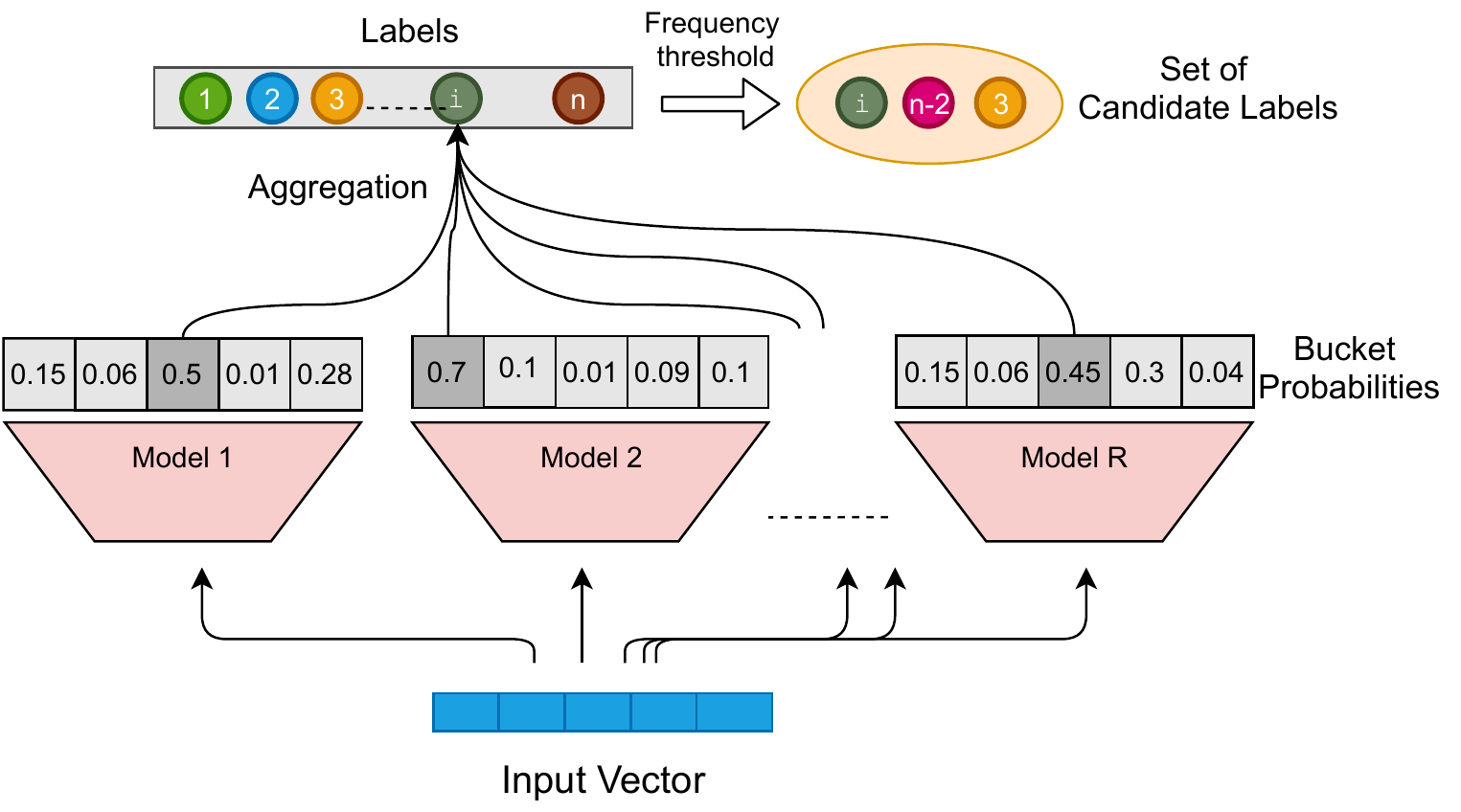}}
\caption{IRLI query process. Here the query vector is passed through $R$ trained models, and each one gives the probability scores over the corresponding buckets. Figure shows $m=1$ for illustration purpose. The top candidates are sorted based on the aggregated scores of each label.}
\label{fig:NeaurlIndexQuery}
\end{center}
\vskip -0.2in
\end{figure}
After training, we store the trained models and inverted indexes for all $R$ repetitions. During the query process, a vector $q \in \mathcal{R}^d$ is passed through $R$ trained nets independently in parallel, where each one gives a $B$ dimensional probability vector. 

We select the top-$m$ buckets ($m$ is $5-10$ if $B$ is around $5000$) from each model. This gives a total of $m \times R$ buckets to probe. A union of points/labels in these buckets is the target candidate set. Additionally, we count each candidate's frequency of occurrence in the total $m \times R$ sets. A higher frequency of a candidate label signifies a higher relevance to the query point. In the end, we keep only the higher relevance candidates by filtering and rejecting the candidates below a certain frequency threshold from the pool.
Please refer to Figure \ref{fig:NeaurlIndexQuery} for an illustrative view of the query process.\\
\textbf{Two crucial points to note}: 1) Our procedure will ensure that every network has $\delta$ higher probability of selecting a relevant label than it can with learning on any predefined random partitioning. We prove this in section \ref{analysis}. 2) Candidate set selection from $R$ repetition and frequency-based filtering exponentially decrease the variance of our true labels estimates. This is analyzed further in Appendix.

\section{Analysis}\label{analysis}

In this section, we theoretically analyze IRLI from two main perspectives. First, we show that the predicted probability of buckets corresponding to the relevant labels increases after re-assigning the labels. Second, analogous to the popular power-of-2-choices, we show that the process of re-assigning to the least occupied of the top-$K$ buckets is the optimal strategy to ensure load balance across the buckets.

As mentioned before, we have $L$ classes being hashed to $B$ buckets using a universal hash function. This randomly partitions the classes into $B$ meta-classes. We estimate the top bucket probability $max\  Pr(b/x)$ for a input vector $x$, where $b \in \{1,2 .. B\}$. Since each of the $R$ repetitions is an instantiation of the same process, we only need an $R$-agnostic proof for the fact that re-assignment enhances the prediction probability of the most relevant bucket.

\begin{theorem}
For a given dataset with $x\in \mathbb{R}^{N \times d}$, and its label $l$, the expected affinity of the input query point $x$ with $l$ increases by a margin of $\delta > 0$ after re-partitioning, i.e.,

$$\mathbb{E}\left(P_{x,h^{'}(l)}^{'}\right) = \mathbb{E}\left(P_{x,h(l)}\right) + \delta$$
% Where $l_i$ is any true label of $x_i$.

where $P_x\in \mathbb{R}^{B}$ is the bucket affinity vector of $x$.  

The increment in the query affinity, results in increment of quality of the retrieved labels. 
\end{theorem}

\textbf{Proof}
Let $x$ be an input vector whose label set is $\bar{y}$. Let $p_{l}$ denote the probability of $l$ being a true label to $x$. 

Let the current partitioning be given by a mapping $h(l)$, where $h(l) \in \{1,2,..,B\}$. Also, assume $l_1, l_2\ \in \bar{y}$ and $l_3 \notin \bar{y}$. 

The affinity score of $x$ for $h(l_1)^{th}$ bucket is given by the summation of probability of label $l_1$ and probability of other labels in the same bucket, i.e.,
$$P_{x,h(l_1)}=p_{l_1}+\sum_{k \neq l_1} \mathbf{1}_{h(k)=h(l_1)} p_{k}$$
$\mathbf{1}$ is the indicator function. 
Now, let us reassign the labels as per section \ref{sec:repartition}. Let the new partition be given by $h^{'} (.)$. If $h(l_1)\neq h(l_2)$ and $h(l_1)= h(l_3)$, we want the re-partitioning to reverse this adversarial scenario, i.e., we expect that

$$h^{'}(l_1) = h^{'}(l_2)\ and\ h^{'}(l_1)\neq h^{'}(l_3)$$

Let $Z$ represent the event of $l_3$ being removed from $l_1$'s bucket and $l_2$ being added to it.

$$P^{'}_{x, h^{'}(l_1)}=p_{l_1}+\sum_{k \neq l_1} \mathbf{1}_{h^{'}(k)=h^{'}(l_1)} p_{k} + \mathbf{1}_Z(p_{l_2}- p_{l_3})$$
Expected affinity is given by-

$$\mathbb{E}\left(P_{x, h^{'}(l_1)}\right)= \mathbb{E}\left(P_{x, h(l_1)}\right) + \mathbb{E}(Z)(p_{l_2}- p_{l_3}) $$
As $l_1$ and $l_2$ are high scoring label of data point $x$, $p_{l_1}=p_{l_2}>p_{l_3}$. Also $\mathbb{E}(Z) > 0 $ when there is a possible reassignment.
and hence $$\mathbb{E}\left(P_{x, h^{'}(l_1)}\right) = \mathbb{E}\left(P_{x, h(l_1)}\right) + \delta$$ where $\delta > 0$.

The main implication of the above theorem is that the bucket containing the relevant label $p_{l_1}$ gets higher aggregated affinity as it will have other true labels with higher probability. It is important to note that this affinity increment is only manifested after retraining on this new partitioning of the labels. The re-partitioning alone does not affect any affinity learned by the neural net.

The increased affinity directly increases the recall during the evaluation. Given $R$ reps, the estimated affinity of $x$ is given by $\hat{P}_{x} = \frac{1}{R} \sum_{r=1}^{R} P^r_{x}$. With increasing $R$ the error in correct label estimation also decreases exponentially.

The same holds for the ANN scenario where the labels are the near true neighbors generated for the IRLI indexing.

\begin{theorem}
Consider the process where at each step, a label is chosen independently and uniformly at random and is inserted into the index. Each new label $l$ inserted in the index chooses $K>K_0$ possible destination bins which are the top-$K$ indices of $P_l$, and is placed in the least full of these bins. For a sufficiently large $t$ , the most crowded bin at time $t$ contains fewer than $\frac{\log(\log(L) + f_1(K))}{\log(K)} + O(1) + f_2(K)$ labels with high probability, where $f_1$ and $f_2$ are monotonically decreasing functions of $K$.
\end{theorem}

\paragraph{Proof:} Proof of this theorem is given in Appendix. It draws parallels from the popular power-of-2-choices framework \cite{mitzenmacher2001power}.

\section{Experiments}
\label{experiments}

\begin{table*}[h]
\resizebox{\linewidth}{!}{
% \scriptsize
\fontsize{7}{9}\selectfont
\begin{tabular}{|c|c|c|c|c|c|c|c|}
\hline
\multicolumn{2}{|c|}{}& \multicolumn{3}{c|}{Wiki-500K} & \multicolumn{3}{c|}{Amazon-670K} \\
\hline
 & Method & P@1 & P@3 & P@5 & P@1 & P@3 & P@5 \\
% \hhline{|=|=|=|=|=|=|=|=|}
\hline
\multirow{4}{*}{Main Baselines} & Neural Indexing (10 buckets) & {\bf 60.77} & {\bf 46.09} & {\bf 43.49} & 35.56 & 32.68 & {\bf 31.02}\\
\cline{2-8}
 & Neural Indexing (5 buckets) & {\bf 60.69} & {\bf 45.78} & {\bf 43.15} & 35.13 & 32.20 & {\bf 30.58}\\
 \cline{2-8}
 & SLICE & 59.89 & 39.89 & 30.12 & {\bf 37.77} & {\bf 33.76} & 30.7 \\
\cline{2-8}
 & Parabel & 59.34 & 39.05 & 29.35 & 33.93 & 30.38 & 27.49 \\
% \hhline{|=|=|=|=|=|=|=|=|}
\hline
\multirow{3}{*}{Other Baselines} & AnnexML & 56.81 & 36.78 & 27.45 & 26.36 & 22.94 & 20.59 \\
\cline{2-8}
 & Pfastre XML & 55.00 & 36.14 & 27.38 & 28.51 & 26.06 & 24.17 \\
\cline{2-8}
 & SLEEC & 30.86 & 20.77 & 15.23 & 18.77 & 16.5 & 14.97 \\

\hline
\end{tabular}
}
\caption{Precision @1, @3, @5 for N-Index on Wiki-500K and Amazon-670K vs popular Extreme Classification benchmarks. }
\label{tab:result_xml}
\end{table*}

\begin{table}[h]
\centering
\begin{tabular}{|c|c|c|}
\hline
Dataset & Wiki-500K & Amz-670K\\
\hline
N-Index (m=10) & 0.56 & 1.08\\
\hline
N-Index (m=5) & \textbf{0.47} & {\bf 0.76}\\
\hline
SLICE & 1.37 & 3.49\\
\hline
Parabel & 2.94 & 2.85\\
\hline
PfastreXML & 6.36 & 19.35\\
\hline
\end{tabular}
\caption{Inference speeds against the fastest Extreme Classification benchmarks.}
\label{tab:time_xml}
\end{table}

\subsection{Multi-label Classification}
\noindent\textbf{Datasets:} We use the dense versions of Wiki-500K and Amazon-670K \cite{jain2019slice} datasets available on the Extreme classification repository \cite{Bhatia16}. The sparse versions of these datasets were scaled down and densified using XML-CNN \cite{xmlCNN} features. Both the datasets have $512$ dimension vectors. Wiki-500K has $501,070$ classes with $1,646,302$ train points and Amazon-670K has $670,091$ classes with $490,449$ train points.\\

\noindent\textbf{Network Parameters:} Each of the $R=32$ models has an input layer of 512 dimensions, a hidden layer of 1024 dimensions, and an output layer of $B=20000$. We train these networks for 30 epochs and re-assign the labels every five epochs.\\

\noindent\textbf{Metrics:} We measure precision at 1,3,5 (denoted by P@1, P@3,P@5). As there are no label vectors provided, we use the corresponding points for each label to re-partition (as explained in section \ref{NeuralIndex}). Here we pay an additional re-partitioning cost of $O(L)$.\\

\noindent\textbf{Hardware and framework:} The experiments were done on a DGX machine with 8
NVIDIA-V100 GPUs. We train with Tensorflow (TF) v1.14 library. We use TF Records data streaming to reduce GPU idle time.\\

\noindent\textbf{Baselines}: We compare IRLI with Parabel \cite{parabel}, SLICE \cite{jain2019slice}, AnnexML \cite{annexml}, Pfast XML \cite{pfastre} and SLEEC \cite{sleec}.\\

\noindent\textbf{Results}: Tables \ref{tab:result_xml} and \ref{tab:time_xml} provide the precision and inference time comparison of 2 IRLI variants ($m=5$ and $m=10$) with all the baselines. We analyze the precision and runtime during inference after selecting the top $m = 5,10$ buckets from each of the $32$ independent indexes. While the best labels are expected to be present in the topmost bucket, we relax this by querying the top 5/10 buckets per index. 
% The frequency counts are more informative with increasing $m$ up to a certain point. This increases precision but decreases the inference speed as more points go through the filter.
We can observe that IRLI Index gives the best precision and runtime, beating all baselines for Wiki-500k dataset. On the Amazon-670K, it is faster and more precise on P@5 metric than the baselines in comparison.

\subsection{Nearest Neighborhood Search}
% In this setup, we have dense vectors for all the items to be indexed. We perform two index creation strategies - one where the entire dataset is used to create the index and the other where we perform data and model parallel ANN search by distributing the index creation and query over 8 GPU nodes. 
\begin{figure}[h!]
  \centering
  \subfigure{\includegraphics[width=\columnwidth]{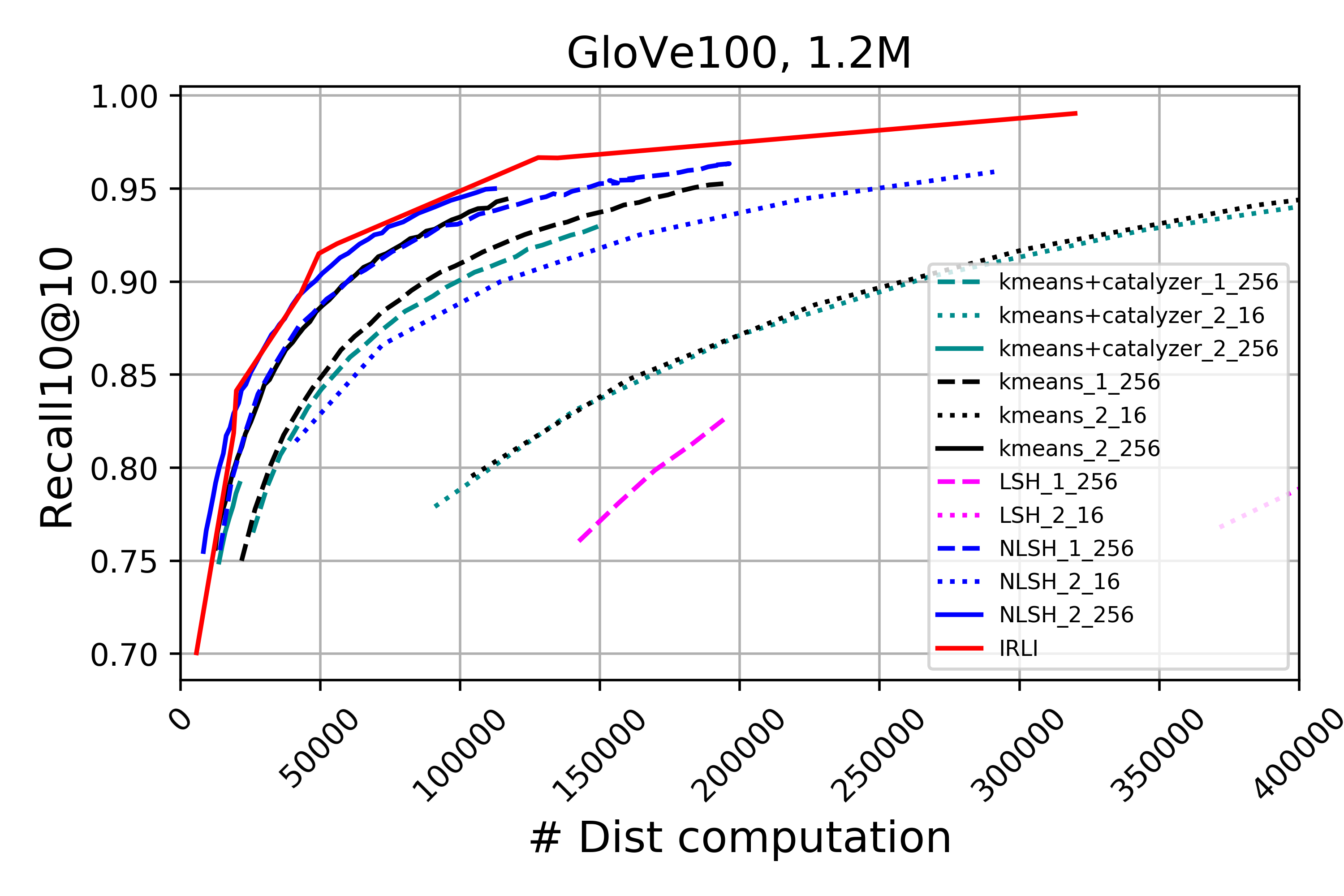}}\\
  \subfigure{\includegraphics[width=\columnwidth]{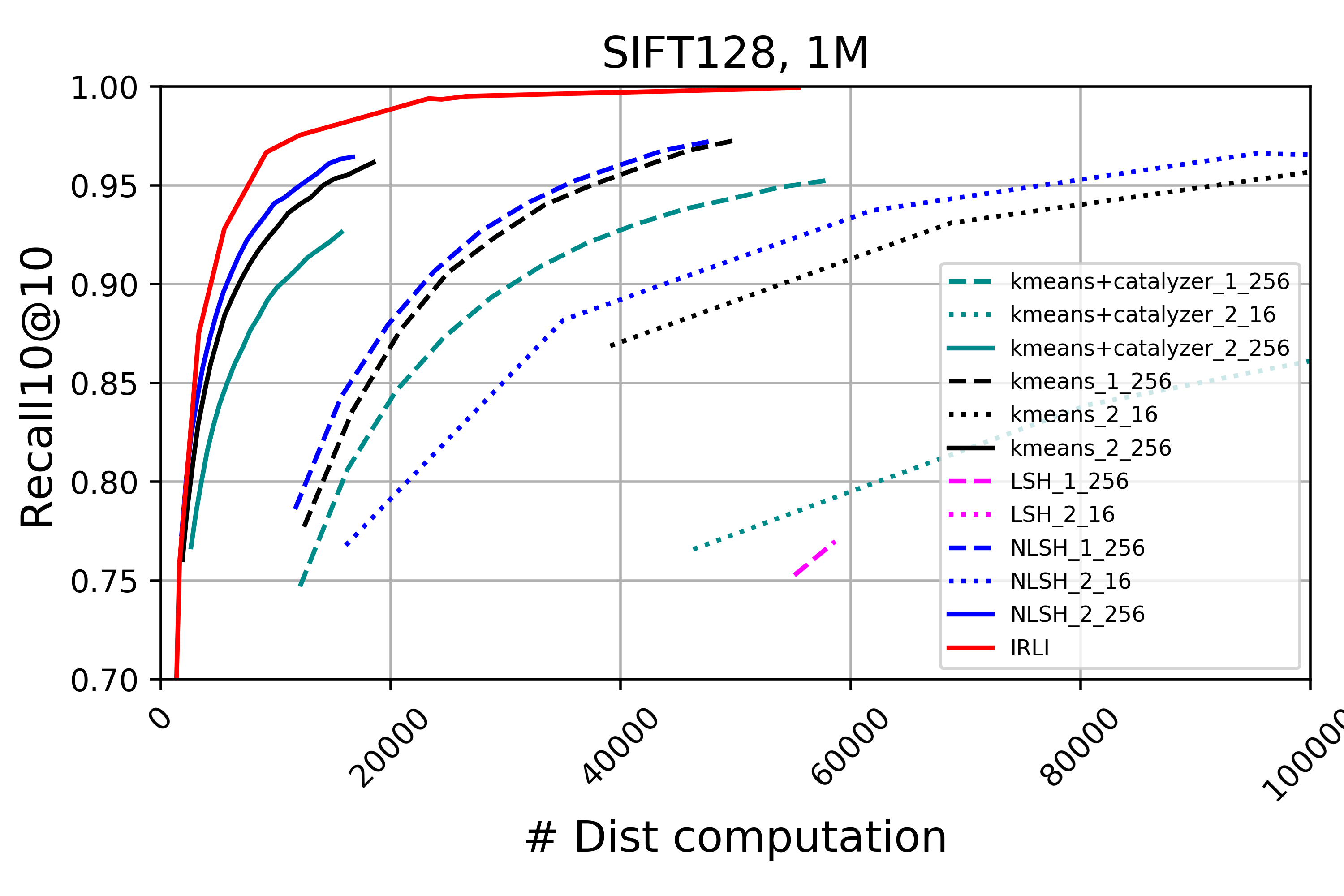}}
  \vspace{-0.4cm}
  \caption{\textbf{Above:} Comparison of IRLI with other partitioning methods on Glove100 dataset: The red curve represents Recall10@10 vs the number of candidates (number of true distance computations) for IRLI. The blue lines represent different variants of the primary baseline NLSH. The best variant of NLSH (2 level-256 bins, 65536 leaf nodes) has no reported recall at larger candidate sizes. All baselines have results at both 1-level and 2-level hierarchies. Balanced k-means, LSH and Catalyzer, are noticeably worse than IRLI and NLSH. \textbf{Below:} Comparison of IRLI with other partitioning methods on SIFT128 dataset. All baselines, including the SOTA NLSH have noticeably worse recall than IRLI.}
  \label{fig:glove_sift_results}
\end{figure}
\noindent\textbf{Dataset:} We have used two million-scale datasets from ANN benchmarks \cite{AnnBenchmark}- Glove100 \cite{glove100_original} and Sift1M. Glove100 has total $1,183,514$ points, each a $100$ dimensional vector and trained with angular distance metric. Sift-1M has exactly $1$ MM points, each a $128$ dimensional vector and trained with the euclidean distance metric.\\

\noindent\textbf{Hyper-parameters:} Each of the $R$ models are simple feed forward networks with an input layer of 100 or 128 (Glove vs SIFT), one hidden layer of 1024 neurons and and output layer of $B=5000$.\\

\noindent\textbf{Hardware and framework:} The experiments were done on a DGX machine with 8 NVIDIA-V100 GPUs. We train with Tensorflow (TF) v1.14 library. We use TF Records data streaming to reduce GPU idle time.\\

\noindent\textbf{Baselines:} We compare IRLI against the popular learned partitioning methods like Neural LSH, kmeans+Neural Catalyzer \cite{NeuralCatalyser} and Cross-Polytype LSH. 
% In addition, we use kmeans with Neural Catalyser \cite{NeuralCatalyser}. It transforms the vectors from the original space to a learned mapping such that it adapts to the index or the quantizer for any discretization process \cite{NeuralCatalyser}. Other baselines are LSH, and Neural LSH \cite{nlsh}.
All baselines are measured with and without hierarchy (1-level and 2-level). Please refer to figure \ref{fig:glove_sift_results} for the detailed comparison plots. For every baseline in the figure, a tag of $1-256$ denotes 1 level, 256 buckets, while a tag of $2-256$ denotes two levels, 256 buckets (65536 leaf nodes effectively).\\

\noindent\textbf{Metrics:} Our metric of interest is the recall of the top 10 neighbors for a particular candidate size. To be precise, both ILRI and the baselines only provide a set of candidate points within which we compute true distance computations to obtain the top 10 closest points. The intersection of this set with the true top-10 neighbors (recall10@10) is the primary metric of our interest. \\
% \begin{figure}[h!]
% \vskip 0.2in
% \begin{center}
% \centerline{\includegraphics[width=\columnwidth]{GloVe100Final.png}}
% \caption{Comparison of IRLI with other partitioning methods on Glove100 dataset. The red curve represents Recall@10 vs number of candidates (number of true distance computations) for IRLI. The blue lines represent different variants of the primary baseline NLSH. The best variant of NLSH (2 level-256 bins, 65536 leaf nodes) has no reported recall at higher candidate sizes. All baselines have results at both 1-level and 2-level hierarchies. Balanced k-means, LSH and Catalyzer are noticeably worse then IRLI and NLSH.}
% \label{fig:glove_result}
% \end{center}
% \vskip -0.2in
% \end{figure}
% \begin{figure}[h!]
% \vskip 0.2in
% \begin{center}
% \centerline{\includegraphics[width=\columnwidth]{SIFT128Final.png}}
% \caption{Comparison of IRLI with other partitioning methods on SIFT128 dataset. All baselines including the SOTA NLSH have noticeably worse recall then IRLI.}
% \label{fig:sift_result}
% \end{center}
% \vskip -0.2in
% \end{figure}
% 

\noindent\textbf{Results} \label{results}
Figure \ref{fig:glove_sift_results} show the comparison with Neural-LSH and other baselines. We can notice that the IRLI (red curve) comfortably surpasses all baselines for a given candidate size. The only baseline configuration that comes close to IRLI is NLSH with a 2-level 256 bin configuration, where the $2^{nd}$ level has 65536 classes and uses k-means clustering instead of a neural network prediction.

\subsubsection{Ablation Study}
\paragraph{Load-Balance:} As mentioned earlier, $K=10$ is an empirical sweet spot for retaining precision while ensuring load-balance during bucket re-assignment. Table \ref{tab:std_vs_k} shows the standard deviation of the bucket load for Glove100 for various values of $K$. We start with a random partition of points (using a 2-universal hash function) and train for 5 epochs, after which we reassign the 1.2 M vectors as explained before. We can observe that as $K$ increases, each bucket has nearly the same number of candidates. However, a larger $K$ might compromise the relevance of buckets. Hence we chose $K=10$ as an appropriate trade-off.

\begin{figure}[h!]
% \vskip 0.2in
\begin{center}
\centerline{\includegraphics[width=\columnwidth]{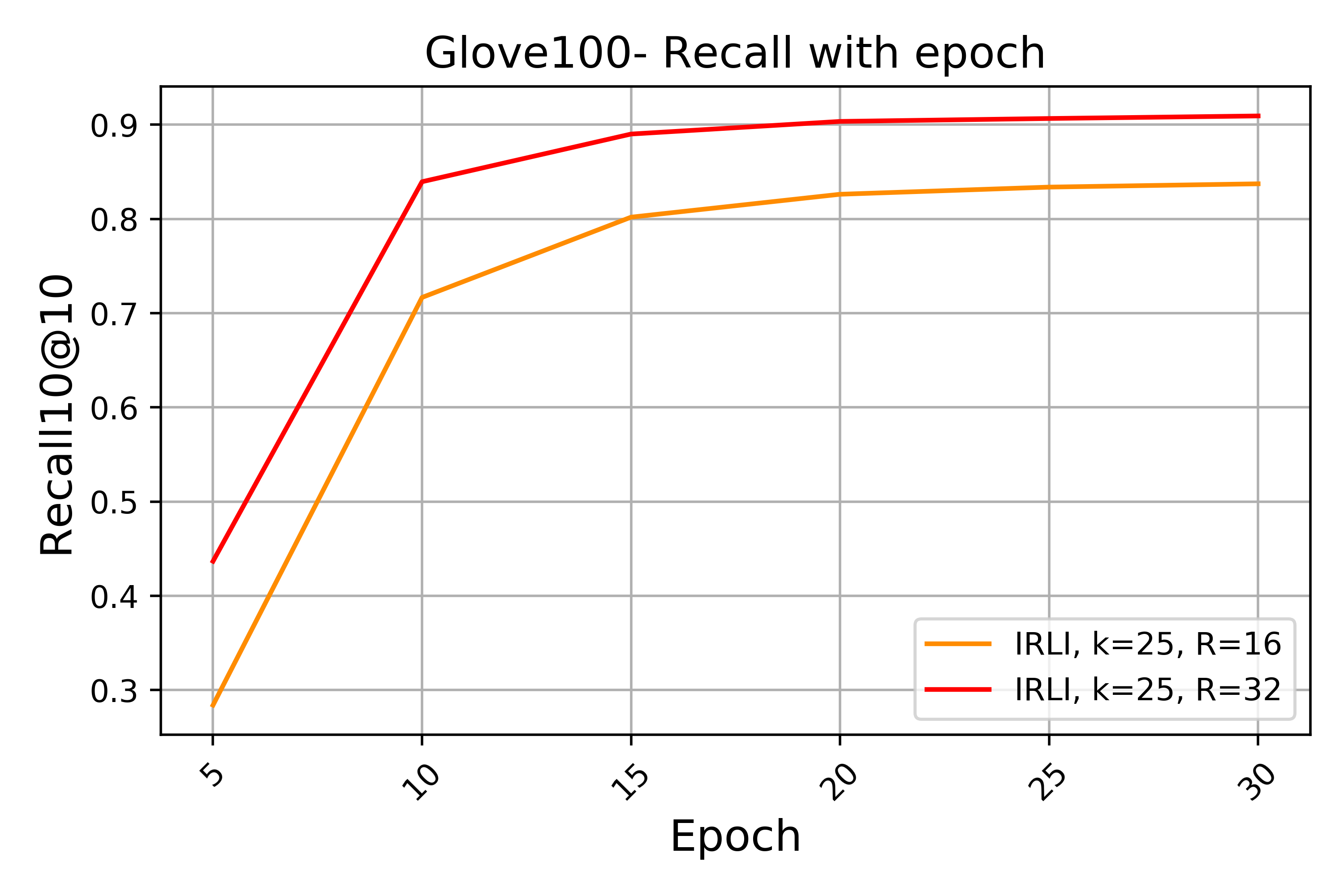}}
\vspace{-0.2cm}
\caption{Epoch-wise R10@10 for Glove100, with $K=25$.}
\label{fig:recallvsEpch}
\end{center}
\vskip -0.2in
\end{figure}

\begin{table}[h!]
\centering
\begin{tabular}{|c|c|c|c|}
\hline
Random Partition & K=5 & K=10 & K=25\\
\hline
15.3 & 17.08 & 2.66 & 0.46\\
\hline
\end{tabular}
\caption{Standard Deviation of load vs $K$ for Glove-100 with $B=5000$ ($mean=236.7$). Larger $K$ gets better load balance while smaller $K$ gets better precision and recall. Although $K=25$ achieves near perfect load balance, $K=10$ is the practical choice for all our experiments for better precision.}
\label{tab:std_vs_k}
\end{table}
\vspace{-5mm}
\paragraph{Epoch-wise Performance:} Figure \ref{fig:recallvsEpch} shows the epoch-wise recall for Glove100 dataset. We can observe that the recall converges after epoch $25$, justifying our choice to train for $30$ epochs. The improvement in recall as we increase $R$ from 16 to 32 is significant. Beyond $R=32$, we do not observe any considerable gain with more repetitions. 

Further, keeping $R=32$ and probing top $m=100$ buckets for a point in each repetition, we measure the number of candidates that appear in at least 4 of the 32 repetitions. As we train more, we expect this candidate set to increase in size as we group more relevant candidates together. Table \ref{tab:cand_size} confirms that trend.

\begin{table}[h!]
\centering
\begin{tabular}{|c|c|c|c|c|c|}
\hline
Epoch-5 & 10 & 15 & 20 & 25 & 30 \\
\hline
5011 & 19151 & 36252 & 43605 & 47072 & 48969 \\
\hline
\end{tabular}
\caption{Candidates appearing atleast 4 of 32 repetitions.}
\label{tab:cand_size}
\end{table}
\vspace{-2mm}
\subsection{Data and Model Distributed KNN on 100M points}
This section demonstrates the scalability of IRLI by partitioning a 100MM subsample of Deep-1B dataset~\cite{deep-1b} and achieving $sub-5\ ms$ latency with a $R10@10$ of $96.16\%$. In the prior cases, each of the $R$ independent models catered to the entire set of labels/data points. In this case, we distribute the 100MM points across $P=8$ disjoint nodes. This choice of $P$ nodes was made to maximize the use of all GPUs on our machine. 

\paragraph{Dataset:} Deep-1B is an image indexing dataset comprising of 1 billion 96-dimensional image descriptors. These descriptors were generated from the last layer of a pre-trained convolutional neural network as discussed in~\cite{deep-1b}. Indexing datasets of this scale is an uphill task and the primary GPU friendly algorithm that accomplishes this is the popular FAISS library~\cite{faiss}.

Deep-1B also provides 350 MM additional training vectors of which we subsample 10 MM vectors to train all models across the 8 nodes.
\paragraph{Hyper-parameters and distribution details:} As mentioned earlier, each of the 8 nodes caters to only 12.5 MM points. For each node, these 12.5 MM points are partitioned into $B=20000$ buckets. Unlike the previous cases ($R=32$), we choose $R=4$ for each node. This would again lead to a total of $P*R\ =\ 32$ models. Each model is a simple feed-forward network with an input dimension of 96 and a hidden layer with 1024 neurons. We train for a total of 20 epochs and reassign the points to the buckets once every 5 epochs.
\paragraph{Hardware and framework:} We use a server equipped with 64 CPU cores and 8 Quadro GPUs each with 48 GB memory. However, our 32 models have a combined parameter set of $660$MM float$32$ values requiring just $2.56$ GB of GPU memory (and 2x auxiliary momentum parameters for Adam optimizer). Like with the previous cases, we train using Tensorflow v1.14.
\paragraph{Results:}\label{results_dist}

\begin{figure}[h!]
\vskip 0.2in
\begin{center}
\centerline{\includegraphics[width=\columnwidth]{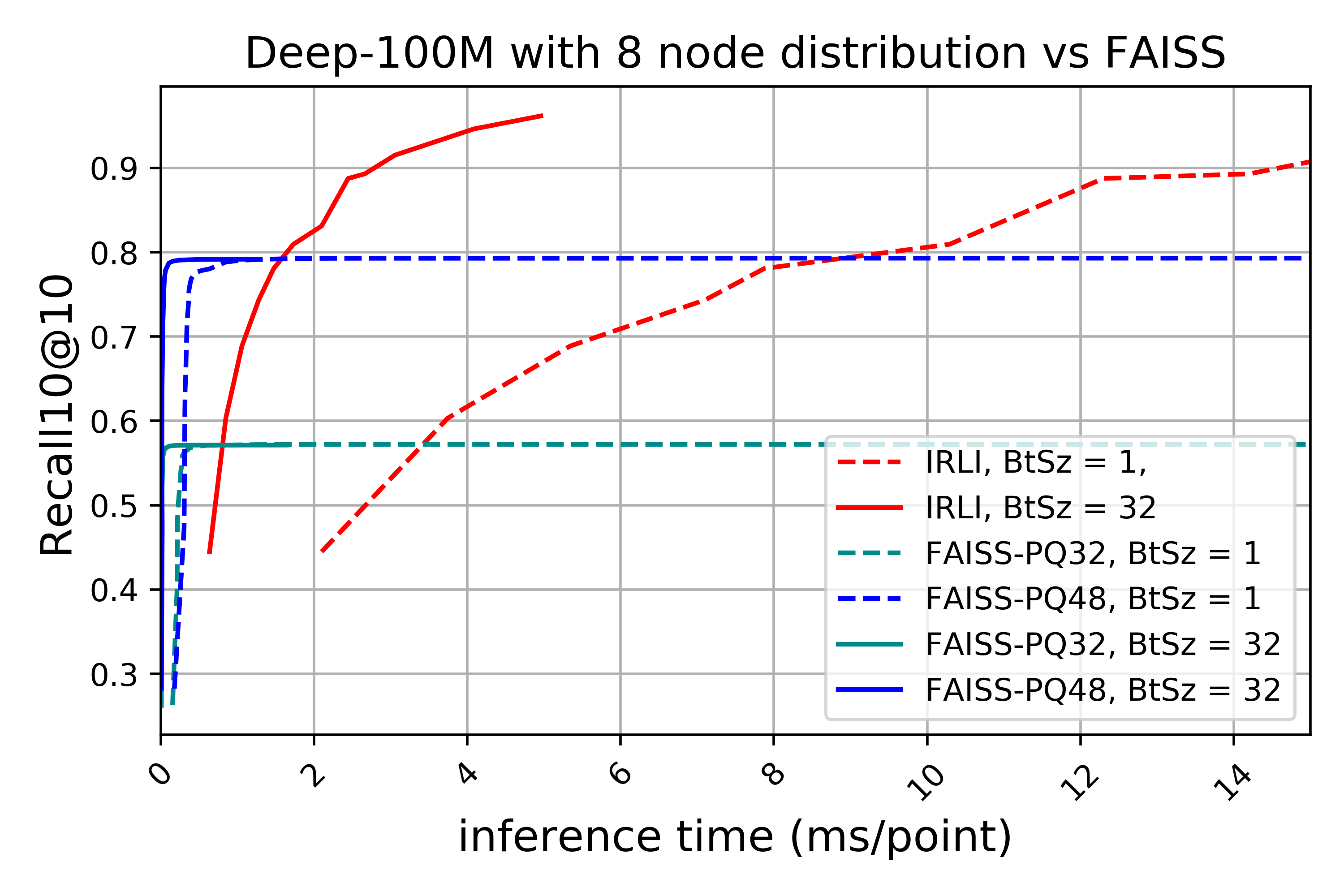}}
\vspace{-0.2cm}
\caption{Recall vs Time comparison of 8-way distributed IRLI (red curve) with FAISS (blue and green) on GPU. BtSz stands for Batch Size. The trade-off of IRLI is governed $m$. We vary $m$ from $1-20$ while keeping $R=4$. The inference time ranges from 0.637 ms (lowest recall) to 4.958 ms (highest recall) per point (solid red curve). FAISS clearly plateaus on recall very soon while IRLI saturate at almost 100\% as we afford more inference time.}
\label{fig:dist_result}
\end{center}
\vskip -0.2in
\end{figure}

\begin{figure}[h!]
\vskip 0.2in
\begin{center}
\centerline{\includegraphics[width=\columnwidth]{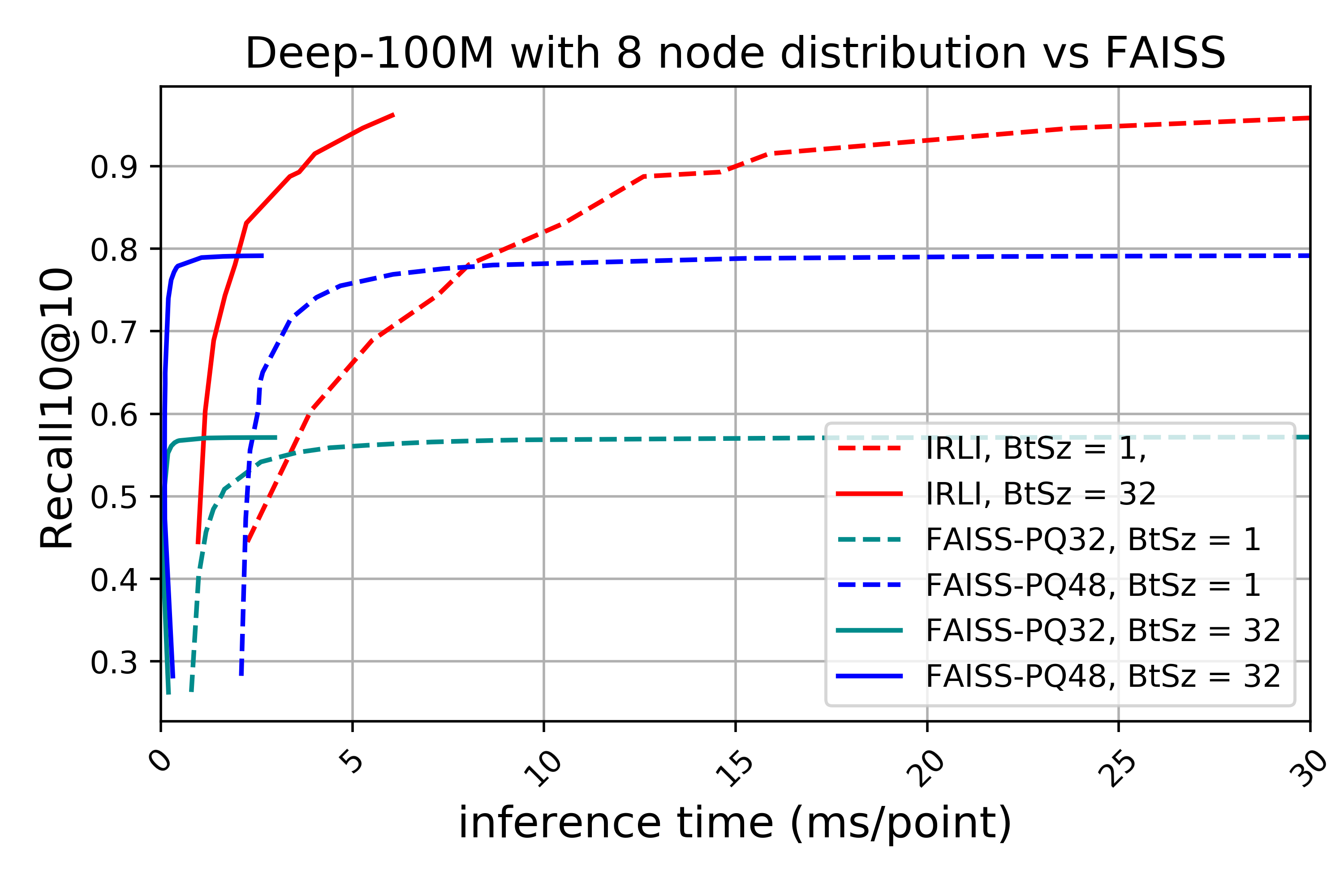}}
\vspace{-0.2cm}
\caption{Recall vs Time comparison of 8-way distributed IRLI (red curve) with FAISS (blue and green) on CPU. As with figure \ref{fig:dist_result}, we vary $m$ from $1-20$ while keeping $R=4$.}
\label{fig:dist_resultCPU}
\end{center}
\vskip -0.2in
\end{figure}

Figures \ref{fig:dist_result} and \ref{fig:dist_resultCPU} compare IRLI against FAISS \cite{faiss} on GPU and CPU respectively (please note the IRLI uses GPU only for the neural network prediction. Rest of the aggregation process is done entirely on CPU). We can observe that FAISS plateaus at 0.8 recall (corroborated by the reported result in \cite{faiss} too). FAISS does product quantization on the 96 dimensions data vectors using $65536$ clusters. In Figure \ref{fig:dist_result}, $PQ32$ refers to a 32 splits of a data vector while $PQ48$ refers to a 48 split. Each vector of the codebook is a byte long. Higher-order splits yield poor recall while $PQ96$ (a quantized version of full product computation) ran out of memory on our GPU. During inference, the recall vs. time trade-off is governed by $m$ (the number of top-scoring buckets that we probe among the $B$). The red curve in figure \ref{fig:dist_result} goes from an inference time of $0.637$ms to $4.958$ms with $m$ ranging from $1$ to $20$. The average candidate set size on which we compute true distances ranges from $19.8K$ to $349K$ (the union of candidates across $8$ nodes).

\bibliographystyle{plainnat}
\bibliography{references}

\begin{thebibliography}{29}
\providecommand{\natexlab}[1]{#1}
\providecommand{\url}[1]{\texttt{#1}}
\expandafter\ifx\csname urlstyle\endcsname\relax
  \providecommand{\doi}[1]{doi: #1}\else
  \providecommand{\doi}{doi: \begingroup \urlstyle{rm}\Url}\fi

\bibitem[Andoni et~al.(2015)Andoni, Indyk, Laarhoven, Razenshteyn, and
  Schmidt]{crossPlotopLSH}
Alexandr Andoni, Piotr Indyk, Thijs Laarhoven, Ilya Razenshteyn, and Ludwig
  Schmidt.
\newblock Practical and optimal lsh for angular distance.
\newblock \emph{arXiv preprint arXiv:1509.02897}, 2015.

\bibitem[Aumüller et~al.(2020)Aumüller, Bernhardsson, and
  Faithfull]{AnnBenchmark}
Martin Aumüller, Erik Bernhardsson, and Alexander Faithfull.
\newblock Ann-benchmarks: A benchmarking tool for approximate nearest neighbor
  algorithms.
\newblock \emph{Information Systems}, 87:\penalty0 101374, 2020.
\newblock ISSN 0306-4379.
\newblock \doi{https://doi.org/10.1016/j.is.2019.02.006}.
\newblock URL
  \url{http://www.sciencedirect.com/science/article/pii/S0306437918303685}.

\bibitem[Babenko and Lempitsky(2016)]{deep-1b}
Artem Babenko and Victor Lempitsky.
\newblock Efficient indexing of billion-scale datasets of deep descriptors.
\newblock In \emph{Proceedings of the IEEE Conference on Computer Vision and
  Pattern Recognition}, pages 2055--2063, 2016.

\bibitem[Bhatia et~al.(2016)Bhatia, Dahiya, Jain, Mittal, Prabhu, and
  Varma]{Bhatia16}
K.~Bhatia, K.~Dahiya, H.~Jain, A.~Mittal, Y.~Prabhu, and M.~Varma.
\newblock The extreme classification repository: Multi-label datasets and code,
  2016.
\newblock URL \url{http://manikvarma.org/downloads/XC/XMLRepository.html}.

\bibitem[Bhatia et~al.(2015)Bhatia, Jain, Kar, Varma, and Jain]{sleec}
Kush Bhatia, Himanshu Jain, Purushottam Kar, Manik Varma, and Prateek Jain.
\newblock Sparse local embeddings for extreme multi-label classification.
\newblock In \emph{Advances in neural information processing systems}, pages
  730--738, 2015.

\bibitem[Dasgupta and Sinha(2013)]{RPTrees}
Sanjoy Dasgupta and Kaushik Sinha.
\newblock Randomized partition trees for exact nearest neighbor search.
\newblock In \emph{Conference on Learning Theory}, pages 317--337. PMLR, 2013.

\bibitem[Dong et~al.(2019)Dong, Indyk, Razenshteyn, and Wagner]{nlsh}
Yihe Dong, Piotr Indyk, Ilya Razenshteyn, and Tal Wagner.
\newblock Learning space partitions for nearest neighbor search.
\newblock \emph{arXiv preprint arXiv:1901.08544}, 2019.

\bibitem[Horvitz and Thompson(1952)]{horvitz1952generalization}
Daniel~G Horvitz and Donovan~J Thompson.
\newblock A generalization of sampling without replacement from a finite
  universe.
\newblock \emph{Journal of the American statistical Association}, 47\penalty0
  (260):\penalty0 663--685, 1952.

\bibitem[Jain et~al.(2016)Jain, Prabhu, and Varma]{pfastre}
Himanshu Jain, Yashoteja Prabhu, and Manik Varma.
\newblock Extreme multi-label loss functions for recommendation, tagging,
  ranking \& other missing label applications.
\newblock In \emph{Proceedings of the 22nd ACM SIGKDD International Conference
  on Knowledge Discovery and Data Mining}, pages 935--944, 2016.

\bibitem[Jain et~al.(2019)Jain, Balasubramanian, Chunduri, and
  Varma]{jain2019slice}
Himanshu Jain, Venkatesh Balasubramanian, Bhanu Chunduri, and Manik Varma.
\newblock Slice: Scalable linear extreme classifiers trained on 100 million
  labels for related searches.
\newblock In \emph{WSDM ’19, February 11–15, 2019, Melbourne, VIC,
  Australia}. ACM, February 2019.
\newblock Best Paper Award at WSDM '19.

\bibitem[Johnson et~al.(2019)Johnson, Douze, and J{\'e}gou]{faiss}
Jeff Johnson, Matthijs Douze, and Herv{\'e} J{\'e}gou.
\newblock Billion-scale similarity search with gpus.
\newblock \emph{IEEE Transactions on Big Data}, 2019.

\bibitem[Khandagale et~al.(2020)Khandagale, Xiao, and Babbar]{bonsai}
Sujay Khandagale, Han Xiao, and Rohit Babbar.
\newblock Bonsai: diverse and shallow trees for extreme multi-label
  classification.
\newblock \emph{Machine Learning}, 109\penalty0 (11):\penalty0 2099--2119,
  2020.

\bibitem[Kraska et~al.(2018)Kraska, Beutel, Chi, Dean, and
  Polyzotis]{kraska2018case}
Tim Kraska, Alex Beutel, Ed~H Chi, Jeffrey Dean, and Neoklis Polyzotis.
\newblock The case for learned index structures.
\newblock In \emph{Proceedings of the 2018 International Conference on
  Management of Data}, pages 489--504, 2018.

\bibitem[Liu et~al.(2017)Liu, Chang, Wu, and Yang]{xmlCNN}
Jingzhou Liu, Wei-Cheng Chang, Yuexin Wu, and Yiming Yang.
\newblock Deep learning for extreme multi-label text classification.
\newblock In \emph{Proceedings of the 40th International ACM SIGIR Conference
  on Research and Development in Information Retrieval}, pages 115--124, 2017.

\bibitem[Lv et~al.(2007)Lv, Josephson, Wang, Charikar, and Li]{multiProbeLSH}
Qin Lv, William Josephson, Zhe Wang, Moses Charikar, and Kai Li.
\newblock Multi-probe lsh: efficient indexing for high-dimensional similarity
  search.
\newblock In \emph{33rd International Conference on Very Large Data Bases, VLDB
  2007}, pages 950--961. Association for Computing Machinery, Inc, 2007.

\bibitem[Malinen and Fr{\"a}nti(2014)]{malinen2014balanced}
Mikko~I Malinen and Pasi Fr{\"a}nti.
\newblock Balanced k-means for clustering.
\newblock In \emph{Joint IAPR International Workshops on Statistical Techniques
  in Pattern Recognition (SPR) and Structural and Syntactic Pattern Recognition
  (SSPR)}, pages 32--41. Springer, 2014.

\bibitem[Malkov and Yashunin(2018)]{malkov2018efficient}
Yu~A Malkov and Dmitry~A Yashunin.
\newblock Efficient and robust approximate nearest neighbor search using
  hierarchical navigable small world graphs.
\newblock \emph{IEEE transactions on pattern analysis and machine
  intelligence}, 42\penalty0 (4):\penalty0 824--836, 2018.

\bibitem[Medini et~al.(2019)Medini, Huang, Wang, Mohan, and Shrivastava]{mach}
Tharun Kumar~Reddy Medini, Qixuan Huang, Yiqiu Wang, Vijai Mohan, and Anshumali
  Shrivastava.
\newblock Extreme classification in log memory using count-min sketch: A case
  study of amazon search with 50m products.
\newblock In \emph{Advances in Neural Information Processing Systems 32}, pages
  13265--13275. 2019.

\bibitem[Mitzenmacher(2001)]{mitzenmacher2001power}
Michael Mitzenmacher.
\newblock The power of two choices in randomized load balancing.
\newblock \emph{IEEE Transactions on Parallel and Distributed Systems},
  12\penalty0 (10):\penalty0 1094--1104, 2001.

\bibitem[Nigam et~al.(2019)Nigam, Song, Mohan, Lakshman, Ding, Shingavi, Teo,
  Gu, and Yin]{dssm}
Priyanka Nigam, Yiwei Song, Vijai Mohan, Vihan Lakshman, Weitian Ding, Ankit
  Shingavi, Choon~Hui Teo, Hao Gu, and Bing Yin.
\newblock Semantic product search.
\newblock In \emph{Proceedings of the 25th ACM SIGKDD International Conference
  on Knowledge Discovery \& Data Mining}, pages 2876--2885, 2019.

\bibitem[Pennington et~al.(2014)Pennington, Socher, and
  Manning]{glove100_original}
Jeffrey Pennington, Richard Socher, and Christopher~D. Manning.
\newblock Glove: Global vectors for word representation.
\newblock In \emph{Empirical Methods in Natural Language Processing (EMNLP)},
  pages 1532--1543, 2014.
\newblock URL \url{http://www.aclweb.org/anthology/D14-1162}.

\bibitem[Prabhu et~al.(2018)Prabhu, Kag, Harsola, Agrawal, and Varma]{parabel}
Yashoteja Prabhu, Anil Kag, Shrutendra Harsola, Rahul Agrawal, and Manik Varma.
\newblock Parabel: Partitioned label trees for extreme classification with
  application to dynamic search advertising.
\newblock In \emph{Proceedings of the 2018 World Wide Web Conference}, pages
  993--1002, 2018.

\bibitem[Sablayrolles et~al.(2018)Sablayrolles, Douze, Schmid, and
  J{\'e}gou]{NeuralCatalyser}
Alexandre Sablayrolles, Matthijs Douze, Cordelia Schmid, and Herv{\'e}
  J{\'e}gou.
\newblock Spreading vectors for similarity search.
\newblock \emph{arXiv preprint arXiv:1806.03198}, 2018.

\bibitem[Sanders and Schulz(2013)]{KaHIP}
Peter Sanders and Christian Schulz.
\newblock Think locally, act globally: Highly balanced graph partitioning.
\newblock In \emph{International Symposium on Experimental Algorithms}, pages
  164--175. Springer, 2013.

\bibitem[Shrivastava and Li(2014)]{shrivastava2014densifying}
Anshumali Shrivastava and Ping Li.
\newblock Densifying one permutation hashing via rotation for fast near
  neighbor search.
\newblock In \emph{International Conference on Machine Learning}, pages
  557--565. PMLR, 2014.

\bibitem[Sproull(1991)]{pcaTrees}
Robert~F Sproull.
\newblock Refinements to nearest-neighbor searching in k-dimensional trees.
\newblock \emph{Algorithmica}, 6\penalty0 (1):\penalty0 579--589, 1991.

\bibitem[Tagami(2017)]{annexml}
Yukihiro Tagami.
\newblock Annexml: Approximate nearest neighbor search for extreme multi-label
  classification.
\newblock In \emph{Proceedings of the 23rd ACM SIGKDD international conference
  on knowledge discovery and data mining}, pages 455--464, 2017.

\bibitem[Wang and Su(2011)]{wang2011improved}
Juntao Wang and Xiaolong Su.
\newblock An improved k-means clustering algorithm.
\newblock In \emph{2011 IEEE 3rd international conference on communication
  software and networks}, pages 44--46. IEEE, 2011.

\bibitem[Wydmuch et~al.(2018)Wydmuch, Jasinska, Kuznetsov, Busa-Fekete, and
  Dembczynski]{extremeText}
Marek Wydmuch, Kalina Jasinska, Mikhail Kuznetsov, R\'{o}bert Busa-Fekete, and
  Krzysztof Dembczynski.
\newblock A no-regret generalization of hierarchical softmax to extreme
  multi-label classification.
\newblock In S.~Bengio, H.~Wallach, H.~Larochelle, K.~Grauman, N.~Cesa-Bianchi,
  and R.~Garnett, editors, \emph{Advances in Neural Information Processing
  Systems}, volume~31, pages 6355--6366. Curran Associates, Inc., 2018.
\newblock URL
  \url{https://proceedings.neurips.cc/paper/2018/file/8b8388180314a337c9aa3c5aa8e2f37a-Paper.pdf}.

\end{thebibliography}

\newpage
.
\newpage
\twocolumn[\icmltitle{Appendix}]
% \section{Proof of Theorem 2}
\begin{lemma}
Consider a process where we sample $K$ elements from a universe of $B$ elements without replacement, with arbitrary probabilities $p_1, p_2,..., p_B$. Denote the score $T$ of this sample set to be the mean of the $K$ sampling probabilities. The variance of $T$ is given by 
$$\frac{1}{K^2}\sum_{i=1}^{K}p_i^2(1-p_i)$$
\end{lemma}

\textbf{Proof:} Suppose we are to measure a characteristic $x_i$ for the $K$ elements in the sample. 

Let us define $T = \sum_{i=1}^K \frac{x_i}{p_i}$. Then, as shown in the theory of importance sampling \cite{horvitz1952generalization}, we have

$$Var(T) = T^2 - \sum_{i=1}^K\frac{x_i^2}{p_i} - \sum_{i\neq j}^K \frac{x_i x_j}{p_ip_j}$$ 

In this theorem, by choosing $x_i = \frac{p_i^2}{K}$, we have $T=\frac{1}{K}\sum_{i=1}^K p_i$ and

$$Var(T) = \frac{(\sum_{i=1}^K p_i)^2}{K^2} - \frac{\sum_{i=1}^K p_i^3}{K^2} - \frac{\sum_{i\neq j}^K p_i p_j}{K^2}$$

$$\implies Var(T) = \frac{1}{K^2}(\sum p_i^2 - \sum p_i^3))$$

$$\implies Var(T) = \frac{1}{K^2}(\sum p_i^2 - \sum p_i^3))$$

It is easy to see that $Var(T)$ is a monotonic decreasing function (say $\textit{f}_m(K)$) of $K$ as it goes down faster than $O(1/K)$.

\textbf{Implications:} Please recall that in IRLI, for any input, we pick the top-$K$ buckets of the $B$ based on the probability scores. Hence, of all the $B$-choose-$K$ combinations, the $K$-tuple that we pick has the maximum mean of probabilities compared to every other $K$-tuple (vice-versa, the $K$-tuple having the maximum mean of probabilities is also the one with all the top-$K$ buckets).

Lemma 1 shows that the variance of the mean of the probabilities of the $K$-tuple decreases with $K$. 

This simulates a virtual random process of picking any tuple with equal likelihood and leads to the following theorem.

\textbf{Theorem 2:} Consider the process where at each step, a label is chosen independently and uniformly at random and is inserted into the index. Each new label $l$ inserted in the index chooses $K>K_0$ possible destination bins which are the top-$K$ indices of $P_l$, and is placed in the least full of these bins. For a sufficiently large $t$ , the most crowded bin at time $t$ contains fewer than $\frac{\log(\log(L) + f_1(K))}{\log(K)} + O(1) + f_2(K)$ labels with high probability, where $f_1$ and $f_2$ are monotonically decreasing functions of $K$.

\paragraph{Proof:} 

Let the number of bins with load $\geq i$  at the end of time $L$ (i.e., the end of the re-partitioning) to be less than $\beta_i$.

Given that we know $\#bins_{\geq i}(L) \leq \beta_i$, we need to find an upper-bound  $\#bins_{\geq i+1}(L)$ to find the max of bins.

In the event of a collision, consider that each label stacks up the on the existing labels like a tower. The height of a label in that case is the number of labels below it. Let $\#labels_{\geq i}(t)$ represent the number of labels that have height $\geq i$ after total $t$ insertions. 

Please note that $\#labels_{\geq i}(t)$ is always higher than the $\#bins_{\geq i}(t)$, as each bin with $\geq i$ has atleast one label with height $\geq i$.

For a new label to land at height $\geq i+1$, all $K$ bins (that we pick) should have load of atleast $i$. With the assumption made in \cite{mitzenmacher2001power}, where the $K$ bins are chosen randomly, the probability of choosing $K$ bins that have height $\geq i$ is at most
$$p_i = \Big(\frac{\beta_i}{B}\Big)^K$$

Since the number of bins with $\frac{2L}{B}$ can atmost be $\frac{B}{2}$, we have $\beta_{\frac{2L}{B}} \leq \frac{B}{2}$.

However, we select the top $K$ bins based on the maximum affinity scores for a given query vector (instead of random $K$ bins). In this case, for any sufficiently large $K \geq K_0$, the probability $p_i$ is atmost
$$p_i \leq \Big(\frac{\beta_i}{B}\Big)^K +\delta$$
where $\delta$ is monotonically decreasing function of $K$ (by invoking \textbf{Lemma 1}).
$$\delta = \textit{f}_m(K)$$ Pardon the abuse of notation, please don't confuse this with the $\delta$ used in Theorem 1.

Hence the $t^{th}$ label has height $\geq i+1$ with probability atmost $p_i$. Number of labels that have height  $\geq i+1$ is atmost $Lp_i$. 
For a fixed IRLI index parameters $L > B$. We can safely assume that
$L = \frac{B}{c}$, where $c <1$.

$$\beta_{i+1} = Lp_i =  \frac{B}{c}\Big(\Big(\frac{\beta_i}{B}\Big)^k +\delta\Big)$$

Just like the case of random $K$ selection, we can set $\beta_{2/c} \leq B/2 +\delta$. 
% As atmost $B/2$ bucket can have balls more than or equal to $2L/B$.\\ 
We now find an expression for $\beta_{2/c +1}$ using induction
$$\beta_{2/c +1} = \frac{B}{c}\Big(\frac{1}{2^K} +\delta_1\Big) $$
$$\beta_{2/c +2} = \frac{B}{c}\Big(\frac{1}{2^{K^2}c^K} +\delta_2\Big) $$

Here each $\delta_i$ is a positive real number, monotonically decreasing in $K$. The $\beta_{2/c +i}$ is given by

$$\beta_{2/c +i} = \frac{B}{c}\Big(\frac{1}{2^{K^i}c^{K^{i-1}}} +\delta_i\Big)$$

Going by the definition of $\beta_i$, if $\beta_i\leq1$, then $i$ represents that maximum load across all the buckets. That happens when 
$$\frac{B}{c}\Big(\frac{1}{2^{K^i}c^{K^{i-1}}} +\delta_i\Big) \leq 1 $$

$$i = \log_K\Big(\log \Big(\frac{L}{1-L\delta_i}\Big)\Big) - \log_K\Big(1+\frac{1}{K}\log c\Big) $$

This can be further simplified into

$$i^* = 2/c +i = \frac{\log \Big(\log L + f_1(K)\Big)}{\log K} + f_2(K) + 2L/B $$

Where $f_1(K) = -\log(1-L\delta_i)$ and $f_2(K) = \log_K(1+\frac{1}{K}\log c)$. As $\delta_i$ decreases with $K$, $f_1(.)$ also monotonically decreases. 

And using the fact that the derivative of $$\frac{log(1+\frac{\log c}{x})}{\log x}$$ is negative for $x>1$, we can conclude that $f_2(.)$ is also a decreasing function of $K$.

Note that, when $K =B$, we are choosing least occupied bin from all $B$ bins. Irrespective of any picking up strategy, $f_1(B)$ will be zero. The bins will be most balanced. With increasing $K$, the load balance increases. However, the chance of label reassignment to a high affinity bin goes down and hence reduces the near-neighbor property of partitions. The value of $K$ used in our experiments is an empirically found optimum.

\end{document}